\documentclass[conference]{IEEEtran}
\IEEEoverridecommandlockouts
\usepackage{cite}
\usepackage{tabularx}
\usepackage{amsmath,amssymb,amsfonts}
\usepackage{algorithm}

\usepackage[noend]{algpseudocode}
\usepackage{graphicx}
\usepackage{textcomp}
\usepackage{xcolor}
\usepackage{float}
\usepackage{booktabs}
\usepackage{array}
\usepackage{wasysym}
\def\BibTeX{{\rm B\kern-.05em{\sc i\kern-.025em b}\kern-.08em
    T\kern-.1667em\lower.7ex\hbox{E}\kern-.125emX}}

\makeatletter
\newcommand{\linebreakand}{%
  \end{@IEEEauthorhalign}
  \hfill\mbox{}\par
  \mbox{}\hfill\begin{@IEEEauthorhalign}
}
\makeatother

\begin{document}

\title{DynTaskMAS: A Dynamic Task Graph-driven Framework for Asynchronous and Parallel LLM-based Multi-Agent Systems\\


\thanks{*These authors contributed equally. This is the preprint version of the conference paper "DynTaskMAS: A Dynamic Task Graph-driven Framework for Asynchronous and Parallel LLM-based Multi-Agent Systems" in \textit{Proc. 35th International Conference on Automated Planning and Scheduling}.}
}

\author{\IEEEauthorblockN{Junwei Yu*}
\IEEEauthorblockA{\textit{The University of Tokyo} \\
Tokyo, Japan \\
yujw@satolab.itc.u-tokyo.ac.jp}
\and
\IEEEauthorblockN{Yepeng Ding*}
\IEEEauthorblockA{\textit{Hiroshima University} \\
Hiroshima, Japan \\
yepengd@acm.org}
\and
\IEEEauthorblockN{Hiroyuki Sato}
\IEEEauthorblockA{\textit{National Institute of Informatics} \\
Tokyo, Japan \\
schuko@nii.ac.jp }
}

\maketitle

\begin{abstract}


The emergence of Large Language Models (LLMs) in Multi-Agent Systems (MAS) has opened new possibilities for artificial intelligence, yet current implementations face significant challenges in resource management, task coordination, and system efficiency. While existing frameworks demonstrate the potential of LLM-based agents in collaborative problem-solving, they often lack sophisticated mechanisms for parallel execution and dynamic task management. This paper introduces DynTaskMAS, a novel framework that orchestrates asynchronous and parallel operations in LLM-based MAS through dynamic task graphs. The framework features four key innovations: (1) a Dynamic Task Graph Generator that intelligently decomposes complex tasks while maintaining logical dependencies, (2) an Asynchronous Parallel Execution Engine that optimizes resource utilization through efficient task scheduling, (3) a Semantic-Aware Context Management System that enables efficient information sharing among agents, and (4) an Adaptive Workflow Manager that dynamically optimizes system performance. Experimental evaluations demonstrate that DynTaskMAS achieves significant improvements over traditional approaches: a 21-33\% reduction in execution time across task complexities (with higher gains for more complex tasks), a 35.4\% improvement in resource utilization (from 65\% to 88\%), and near-linear throughput scaling up to 16 concurrent agents (3.47× improvement for 4× agents). Our framework establishes a foundation for building scalable, high-performance LLM-based multi-agent systems capable of handling complex, dynamic tasks efficiently.

\end{abstract}


\section{Introduction}
The rapid advancement of LLMs has revolutionized the landscape of artificial intelligence, demonstrating unprecedented capabilities in natural language understanding and generation\cite{brown2020language}. These models have shown remarkable proficiency in tasks ranging from text completion to complex reasoning, opening new avenues for intelligent system design. Concurrently, the concept of MAS has gained significant traction, offering a distributed approach to problem-solving that leverages the collective intelligence of multiple entities\cite{liu2023agentbenchevaluatingllmsagents,khattab2023dspy,agentscope,park2023generative,li2023camel,yang2023auto}. The convergence of these two paradigms presents a promising frontier in AI research, with the potential to address increasingly complex and dynamic challenges across various domains

LLM-driven multi-agent systems have exhibited substantial potential in tackling intricate problems that require diverse expertise and collaborative reasoning. However, current research in this field has predominantly focused on simplified architectures with single topology structures and straightforward task flows. While these approaches have yielded valuable insights, they fall short in addressing the multifaceted demands of real-world applications, where task complexity and environmental dynamics can vary significantly.

As the complexity of tasks escalates, traditional MAS architectures face several critical challenges. Firstly, the task decomposition process becomes increasingly intricate, requiring more sophisticated mechanisms to break down complex problems into manageable subtasks while maintaining logical coherence. Secondly, the parallel processing capabilities of existing systems are often underutilized, leading to inefficiencies in resource allocation and execution time. Thirdly, context management across multiple agents becomes exponentially more challenging as the number of interactions and the volume of shared information grow.


To address these limitations, we propose DynTaskMAS, a Dynamic Task Graph-driven Framework for Asynchronous and Parallel LLM-based Multi-Agent Systems. By placing a dynamic task graph, DynTaskMAS enables flexible task decomposition and efficient parallel execution, effectively overcoming the aforementioned challenges. The key contributions of this work are as follows:



\begin{figure*}[ht]
\centerline{\includegraphics[width=0.97\linewidth]{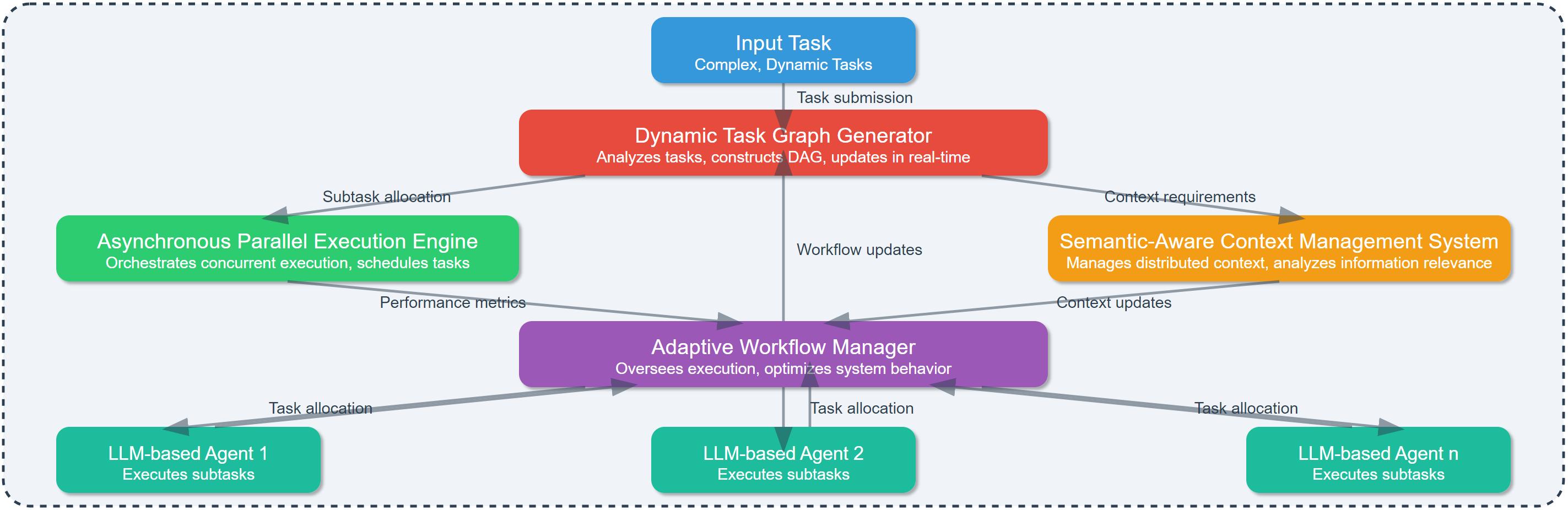}}
\caption{\textbf{The Overview of DynTaskMAS Framework.} The diagram depicts the architecture of the DynTaskMAS framework as a multi-layered structure characterized by bidirectional information flow. At the top, the "Input Task" node serves as the entry point for complex tasks. Below it, DTGG acts as a core component, continuously analyzing and updating task graphs, with arrows connecting to the input task to signify this iterative process. The third layer comprises two parallel modules: APEE and SACMS. Both are linked to DTGG above and AWM below, reflecting their interdependent functions. Positioned in the fourth layer, AWM maintains bidirectional connections with all upper components, underscoring its role in managing and optimizing workflows. At the base, multiple "LLM-based Agent" nodes are connected to APEE and SACMS, illustrating the distributed execution of tasks and continuous exchange of information.}
\label{fig:overview}
\end{figure*}

\begin{enumerate}
\item This work proposes a novel framework for dynamic task orchestration in LLM-based MAS. The approach introduces new principles for decomposing complex language tasks while maintaining semantic coherence and causal dependencies.
\item The research advances the understanding of parallel execution patterns in LLM-based systems through new abstractions for managing asynchronous agent interactions. This theoretical foundation enables more efficient resource utilization while preserving the complex reasoning capabilities inherent to language models.
\item A systematic approach to context management in multi-agent LLM systems addresses the fundamental challenge of maintaining semantic relevance across distributed agents. The methodology introduces new ways to balance information sharing with computational efficiency, establishing principles for scalable LLM-based architectures.
\item Experimental results demonstrate substantial improvements over traditional approaches: 21-33\% reduction in execution time, 35.4\% increase in resource utilization, and near-linear throughput scaling up to 16 concurrent agents.
\end{enumerate}

\section{Background}

\subsection{Planning with LLMs}

Recent studies have explored the capacity of LLMs to perform zero-shot planning, converting high-level natural language tasks into actionable steps without additional training, albeit with noted challenges in precision mapping to executable actions\cite{huang2022language}. Other research has focused on enhancing LLMs' reasoning capabilities through structured prompting techniques, eliciting more deliberate and logical reasoning processes\cite{fagbohun2024empirical}. The integration of LLMs with robotic affordances has been demonstrated to enable complex instruction following in real-world scenarios, showcasing the potential for LLMs to ground language in physical interactions\cite{li2024embodied}. Further advancements have been made in few-shot grounded planning, where LLMs are employed to generate plans for embodied agents within visual environments. Interactive planning methods have been developed to facilitate open-world multi-tasking, leveraging LLMs to describe, explain, plan, and select actions\cite{rasal2024navigating}. Additionally, frameworks that empower LLMs with optimal planning proficiency have been proposed, aiming to solve planning problems articulated in natural language\cite{xiong2024deliberate}. Problem-solving and decision-making capabilities of LLMs have been augmented through novel frameworks designed to mimic human strategic planning\cite{yao2024tree}. Collectively, these contributions underscore the evolving role of LLMs in planning and reasoning, highlighting their potential to enhance embodied agent performance across a spectrum of tasks.

Building upon these contributions, which have significantly advanced the understanding of LLMs in planning and reasoning, there is a notable gap in research focusing on the operational efficiency and practical considerations of resource allocation, particularly in the context of real-time GPU resource management. While these studies have underscored the evolving role of LLMs in enhancing embodied agent performance, they have largely overlooked the challenges associated with the asynchronous and parallel execution of tasks within multi-agent systems. 

\subsection{Multi-Agent System}

In the field of multi-agent systems utilizing large language models, several notable approaches have emerged. AutoGPT demonstrates an autonomous system where agents collaboratively achieve goals through iterative planning, execution, and evaluation cycles\cite{yang2023auto}. Building on this concept, MetaGPT emulates a startup team structure, employing role-specific agents to tackle complex tasks like software development and project planning\cite{hong2023metagptmetaprogrammingmultiagent}. Camel (Collaborative Agents for Multi-agent Environment Learning) further enhances collaborative problem-solving by simulating scenarios that require communication, knowledge sharing, and collective decision-making among language models\cite{li2023camel}. Stanford University's Generative Agents take a different approach, focusing on simulating lifelike behavior in digital environments by creating personas with memory systems and adaptive behaviors\cite{park2023generative}. While these multi-agent systems showcase innovative approaches to collaboration and task execution, they have not yet provided superior solutions for parallel and serial processing challenges, indicating an area for potential future research and development in the field.

\section{DynTaskMAS}

DynTaskMAS is a novel framework designed to enhance the efficiency and adaptability of LLM-based multi-agent systems in handling complex, dynamic tasks. The architecture comprises four primary components that work in concert to achieve flexible task management and optimized resource utilization:

\noindent\textbf{Dynamic Task Graph Generator (DTGG):} This component analyzes incoming tasks and automatically constructs a directed acyclic graph (DAG) representing subtasks and their interdependencies. The DTGG continuously updates the graph as new information becomes available or task requirements change, ensuring adaptability to dynamic environments.

\noindent\textbf{Asynchronous Parallel Execution Engine (APEE):} The APEE orchestrates the concurrent execution of subtasks across multiple LLM-based agents. It employs sophisticated scheduling algorithms to maximize parallelism while respecting task dependencies defined in the dynamic task graph.

\noindent\textbf{Semantic-Aware Context Management System (SACMS):} This subsystem facilitates efficient information sharing among agents by maintaining a hierarchical, distributed context repository. The SACMS employs semantic analysis to determine the relevance of information, ensuring agents have access to pertinent data without unnecessary overhead.

\noindent\textbf{Adaptive Workflow Manager (AWM):} The AWM oversees the overall execution process, dynamically adjusting workflows based on real-time performance metrics and environmental changes. It interfaces with all other components to optimize system behavior and resource allocation.

\begin{figure*}[htbp]
\centerline{\includegraphics[width=0.99\linewidth]{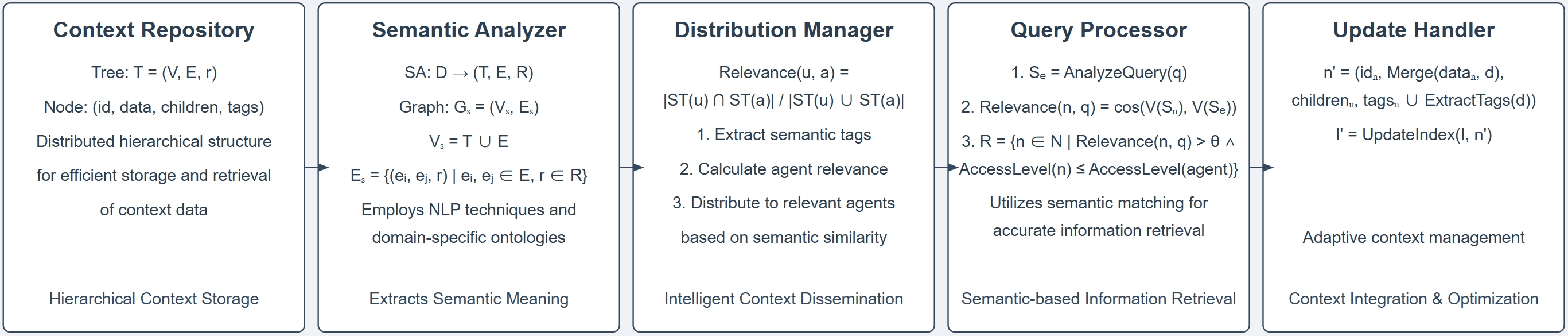}}
\caption{\textbf{The architecture of Semantic-Aware Context Management System.} The SACMS architecture comprises five core components: (1) the Context Repository, a distributed, hierarchical data store designed for efficient storage and retrieval of contextual information; (2) the Semantic Analyzer, which employs advanced natural language processing and domain-specific ontologies to extract meaningful semantic tags and relationships; (3) the Context Distribution Manager, responsible for the efficient dissemination of relevant contextual information to agents based on task requirements and semantic relevance; (4) the Query Processor, which utilizes semantic matching to process context retrieval requests and deliver the most pertinent information; and (5) the Update Handler, which integrates new or updated context data into the repository while maintaining the semantic index. Together, these components enable intelligent, context-aware task execution within the DynTaskMAS system.}
\label{fig:sacms}
\end{figure*}

In the illustrated Figure \ref{fig:overview}, these components interact through a central coordination mechanism that ensures coherent system operation. The modular design of DynTaskMAS allows for scalability and easy integration of additional agents or task types, making it adaptable to various application domains.

\subsection{Dynamic Task Graph Generator}

The Dynamic Task Graph Generator is a crucial component of DynTaskMAS, responsible for decomposing complex tasks into manageable subtasks and representing their dependencies as a DAG. The DTGG continuously updates this graph based on new information and changing task requirements\cite{kwok1999benchmarking}.

\noindent\textbf{Graph Structural Composition:} Let $T = {t_1, t_2, ..., t_n}$ be the set of all tasks in the system. The dynamic task graph $G$ is defined as:
\begin{equation}
G = (V, E, W)
\end{equation}
where \( V = \{ v_1, v_2, \dots, v_m \} \) is the set of vertices representing subtasks, \( E \subseteq V \times V \) is the set of edges indicating dependencies, and \( W: E \rightarrow \mathbb{R}^+ \) is a weight function assigning positive real values to the edges.


The DTGG employs a recursive decomposition algorithm to break down complex tasks. For each task $t_i \in T$, we define a decomposition function $D$:
\begin{equation}
D(t_i) = {s_{i1}, s_{i2}, ..., s_{ik}}
\end{equation}
where $s_{ij}$ are subtasks of $t_i$. The decomposition continues until a predefined granularity level is reached.

The weight of an edge $(v_i, v_j) \in E$ is calculated based on the estimated computational complexity and data dependency between subtasks:
\begin{equation}
W(v_i, v_j) = \alpha \cdot C(v_j) + \beta \cdot I(v_i, v_j)
\end{equation}
where \( C(v_j) \) denotes the estimated computational complexity of subtask \( v_j \), \( I(v_i, v_j) \) represents the context transfer time managed by SACMS from \( v_i \) to \( v_j \), and \( \alpha, \beta \) are balancing coefficients.
For optimal parameter selection, we recommend:
\begin{itemize}
    \item \( \alpha = \frac{1}{T_c} \), where \( T_c \) is the average computation time per complexity unit
    \item \( \beta = \frac{1}{T_t} \), where \( T_t \) is the average context transfer time per unit of information
\end{itemize}
This normalization ensures that both computational complexity and context transfer time contribute proportionally to the edge weight, enabling more accurate task scheduling decisions. The ratio \( \frac{\alpha}{\beta} \) should be adjusted based on the system's relative speeds of computation versus context transfer, typically ranging from 0.5 to 2.0 depending on the specific hardware configuration and network conditions.


\noindent\textbf{Control of Cyclic Dependencies:} To manage the complexity of reflection cycles in the DTGG, it is essential to distinguish between true cyclic dependencies and apparent cyclic structures. In LLM-based MAS, where each prompt functions as an agent, genuine cyclic dependencies primarily manifest in reflection processes, where an agent evaluates and refines its output. Other apparent cycles, such as iterative refinement or progressive enhancement, are essentially linear task sequences with clear hierarchical dependencies. 

To prevent infinite loops or excessive processing in reflection cycles, we implement a maximum iteration threshold \( N \) (typically \( N \leq 3 \)). The reflection terminates when either:
\begin{itemize}
    \item The quality assessment meets the predetermined threshold
    \item The number of reflection cycles reaches \( N \)
    \item The improvement between successive iterations falls below a minimum threshold \( \varepsilon \)
\end{itemize}



The DTGG continuously updates the task graph based on new information and task progress. Let $G_t$ be the graph at time $t$, and $\Delta_t$ be the set of changes at time $t$. The update function $U$ is defined as:
\begin{equation}
G_{t+1} = U(G_t, \Delta_t)
\end{equation}


\begin{algorithm}
\caption{Dynamic Task Graph Generator}
\label{dtgg}
\begin{algorithmic}[1]
\Function{UpdateTaskGraph}{$G, T_{new}, \Delta$}
\For{each $t_i$ in $T_{new}$}
\State $S_i \gets \Call{DecomposeTask}{t_i}$
\State $V \gets V \cup S_i$
\For{each $s_{ij}, s_{ik}$ in $S_i$ where $j < k$}
\State $E \gets E \cup (s_{ij}, s_{ik})$
\State $W(s_{ij}, s_{ik}) \gets \Call{CalculateWeight}{s_{ij}, s_{ik}}$
\EndFor
\EndFor
\For{each change $\delta$ in $\Delta$}
\State $G \gets \Call{ApplyChange}{G, \delta}$
\EndFor
\State \Return $G$
\EndFunction
\Function{DecomposeTask}{$t$}
\If{$\Call{IsAtomicTask}{t}$}
\State \Return ${t}$
\Else
\State $S \gets \emptyset$
\For{each subtask $s$ of $t$}
\State $S \gets S \cup \Call{DecomposeTask}{s}$
\EndFor
\State \Return $S$
\EndIf
\EndFunction
\Function{CalculateWeight}{$v_i, v_j$}
\State $C_j \gets \Call{EstimateComplexity}{v_j}$
\State $I_{ij} \gets \Call{EstimateInformationTransfer}{v_i, v_j}$
\State \Return $\alpha \cdot C_j + \beta \cdot I_{ij}$
\EndFunction
\end{algorithmic}
\end{algorithm}


The pseudocode of the main DTGG algorithm is provided in Algorithm \ref{dtgg}. DTGG Receives complex tasks and updates from the system's input layer. The DTGG's output, a continuously updated task graph, serves as the foundation for efficient task distribution and execution in the DynTaskMAS framework. Its ability to dynamically adjust to changing conditions ensures the system's adaptability in complex, evolving environments.

\subsection{Asynchronous Parallel Execution Engine}
The Asynchronous Parallel Execution Engine is a critical component of DynTaskMAS, responsible for efficiently scheduling and executing tasks across multiple LLM-based agents. It leverages the dynamic task graph generated by the DTGG to maximize parallelism while respecting task dependencies. The APEE consists of the following sub-components: Task Scheduler, Execution Queue Manager, Agent Pool Manager, Load Balancer and Asynchronous Communication Handler.

\noindent\textbf{Task Scheduler:} The Task Scheduler is responsible for determining the execution order of tasks based on the dynamic task graph. It employs a priority-based scheduling algorithm that considers task dependencies, estimated execution time, and system load.
Let $G = (V, E, W)$ be the current task graph. For each task $v_i \in V$, we define its priority $P(v_i)$ as:
\begin{equation}
P(v_i) = \frac{C(v_i)}{\max_{v_j \in \text{Succ}(v_i)} (W(v_i, v_j) + P(v_j))}
\end{equation}
where $C(v_i)$ is the estimated computational complexity of task $v_i$, $\text{Succ}(v_i)$ is the set of immediate successors of $v_i$ in the graph, and $W(v_i, v_j)$ is the weight of the edge $(v_i, v_j)$.

\noindent\textbf{Execution Queue Manager:} The Execution Queue Manager maintains a priority queue of ready-to-execute tasks. It continuously updates the queue based on the Task Scheduler's output and the current system state.
\begin{algorithm}
\caption{Execution Queue Manager}
\begin{algorithmic}[1]
\Function{UpdateExecutionQueue}{$G, Q$}
\State $R \gets {v \in V : \text{Pred}(v) = \emptyset}$ \Comment{Ready tasks}
\For{each $v \in R$}
\State $priority \gets \Call{CalculatePriority}{v, G}$
\State $Q.\Call{Enqueue}{v, priority}$
\EndFor
\State \Return $Q$
\EndFunction
\Function{CalculatePriority}{$v, G$}
\If{$\text{Succ}(v) = \emptyset$}
\State \Return $C(v)$
\Else
\State \Return $\frac{C(v)}{\max_{u \in \text{Succ}(v)} (W(v, u) + \Call{CalculatePriority}{u, G})}$
\EndIf
\EndFunction
\end{algorithmic}
\end{algorithm}

\noindent\textbf{Agent Pool Manager and Load Balancer:} The Agent Pool Manager and Load Balancer work together to efficiently manage and distribute tasks among LLM-based agents. The Agent Pool Manager oversees the pool of available agents, tracking their status, capabilities, and workload. It provides the necessary interface for the Load Balancer, which ensures optimal task distribution based on task priorities, agent capabilities, and current system load. Together, they coordinate to allocate tasks to the most suitable agents, maintaining efficient system performance.





\noindent\textbf{Asynchronous Communication Handler:} The Asynchronous Communication Handler manages the non-blocking communication between the APEE and the LLM-based agents. It uses an event-driven architecture (task completion notifications, agent availability updates, and task failure reports) to handle task assignments, status updates, and result collections to ensure high throughput and responsiveness.

By leveraging the dynamic task graph and employing sophisticated scheduling and load balancing algorithms, the APEE enables DynTaskMAS to achieve high levels of parallelism and efficiency in executing complex, interdependent tasks across multiple LLM-based agents.

\subsection{Semantic-Aware Context Management System}

The SACMS is a pivotal component of DynTaskMAS, designed to efficiently manage and distribute contextual information among LLM-based agents. Figure \ref{fig:sacms} depicts the architecture and key algorithms of SACMS, highlighting its role in enabling intelligent and context-aware task execution. By leveraging advanced semantic analysis techniques, distributed architecture, and adaptive mechanisms, SACMS enables LLM-based agents to access and utilize contextual information effectively, thereby enhancing the overall performance and adaptability of the system.


\noindent\textbf{Context Repository:} The Context Repository serves as the central data store for contextual information. It is implemented as a distributed, hierarchical structure to facilitate efficient storage and retrieval of context data.

The repository is organized as a forest of trees, where each tree represents a distinct context domain. This structure allows for natural representation of hierarchical relationships within contexts.
\begin{equation}
\text{ContextForest} = {T_1, T_2, ..., T_n}
\end{equation}
where each $T_i$ is a context tree defined as:
\begin{equation}
T_i = (V_i, E_i, r_i)
\end{equation}
with $V_i$ being the set of nodes, $E_i$ the set of edges, and $r_i$ the root node of the tree.

Each node in the context tree is defined as follows:
\begin{equation}
\text{ContextNode} = (id, data, children, semanticTags)
\end{equation}
where $id$ is a unique identifier, $data$ contains the actual context information, $children$ is a set of child nodes, and $semanticTags$ is a set of semantic labels associated with the node.

\noindent\textbf{Semantic Analyzer:} The Semantic Analyzer is responsible for processing contextual information to extract meaningful semantic tags and relationships. It employs specialized LLMs and domain-specific ontologies to achieve this.


The semantic analysis process can be formalized as follows:
\begin{equation}
SA: D \rightarrow (T, E, R)
\end{equation}
where $D$ is the input context data, $T$ is the set of extracted tags, $E$ is the set of identified entities, and $R$ is the set of relationships between entities.


The analyzer constructs a semantic graph $G_s = (V_s, E_s)$ where:
\begin{equation}
V_s = T \cup E
\end{equation}
\begin{equation}
E_s = {(e_i, e_j, r) | e_i, e_j \in E, r \in R}
\end{equation}
This graph representation enables efficient semantic querying and reasoning over the context data.

\noindent\textbf{Context Distribution Manager:} The Context Distribution Manager ensures that relevant contextual information is efficiently disseminated to the appropriate agents based on their current tasks and semantic relevance.

The relevance of a context update to an agent is computed using a semantic similarity measure:
\begin{equation}
\text{Relevance}(u, a) = \frac{|ST(u) \cap ST(a)|}{|ST(u) \cup ST(a)|}
\end{equation}
where $ST(u)$ and $ST(a)$ are the sets of semantic tags associated with the update and the agent's current context, respectively.


The distribution process follows Algorithm \ref{alg:context_distribution}.

\begin{algorithm}[t]
\caption{Context Distribution}
\begin{algorithmic}[1]
\Function{DistributeContext}{$update$, $agents$}
    \State $R \gets \emptyset$  \Comment{Relevant agents set}
    \State $tags \gets \Call{ExtractSemanticTags}{update}$
    \For{$agent \in agents$}
        \State $agentTags \gets \Call{GetTags}{agent}$
        \If{$\Call{Relevance}{tags, agentTags} > \theta$}
            \State $R \gets R \cup \{agent\}$
        \EndIf
    \EndFor
    \For{$agent \in R$}
        \State \Call{SendUpdate}{$agent$, $update$}
    \EndFor
\EndFunction
\label{alg:context_distribution}
\end{algorithmic}
\end{algorithm}

%

\noindent\textbf{Query Processor} The Query Processor handles context retrieval requests from agents, utilizing semantic matching to return the most relevant information.


Given a query $q$, we extract its semantic representation:
\begin{equation}
S_q = \text{AnalyzeQuery}(q)
\end{equation}
This transformation ensures consistent interpretation of queries across the distributed system while preserving their semantic intent.

Relevance Assessment employs cosine similarity between vectorized semantics. The relevance of a context node $n$ to a query $q$ is computed as:
\begin{equation}
\text{Relevance}(n, q) = \cos(V(S_n), V(S_q))
\end{equation}
where $V(S_n)$ and $V(S_q)$ are vector representations of the node's and query's semantics, respectively.

The final set of results $R$ is obtained by applying an access control filter:
\begin{equation}
\begin{split}
R = \{n \in N \mid & \text{Relevance}(n, q) > \theta \;\land \\
& \text{AccessLevel}(n) \leq \text{AccessLevel}(agent)\}
\end{split}
\end{equation}
where $\theta$ is a relevance threshold and $N$ is the set of all context nodes. Controlled Filtering applies a dual-constraint mechanism that ensures both relevance optimization and security compliance in result generation.

\noindent\textbf{Update Handler:} Then, the Update Handler manages the process of integrating new or updated context information into the repository, which uses a two-phase commit protocol to maintain context consistency.

Integration Phase performs atomic updates while preserving semantic relationships. The update process for a node $n$ with new data $d$ is defined as:

\begin{equation}
\begin{split}
n' = (&id_n, \text{Merge}(data_n, d), children_n,\\
      &semanticTags_n \cup \text{ExtractTags}(d))
\end{split}
\end{equation}

After each update, the semantic index is updated to reflect the new information:
\begin{equation}
I' = \text{UpdateIndex}(I, n')
\end{equation}
where $I$ is the current semantic index and $I'$ is the updated index. Index Synchronization ensures query consistency post-update. This operation maintains the system's ACID properties while minimizing query latency.





The Semantic-Aware Context Management System provides a robust and efficient solution for context management in DynTaskMAS. By leveraging semantic analysis capabilities from LLMs, SACMS enables LLM-based agents to access and utilize contextual information effectively, thereby enhancing the overall performance and adaptability of the system.

\begin{figure*}[htbp]
\centerline{\includegraphics[width=0.99\linewidth]{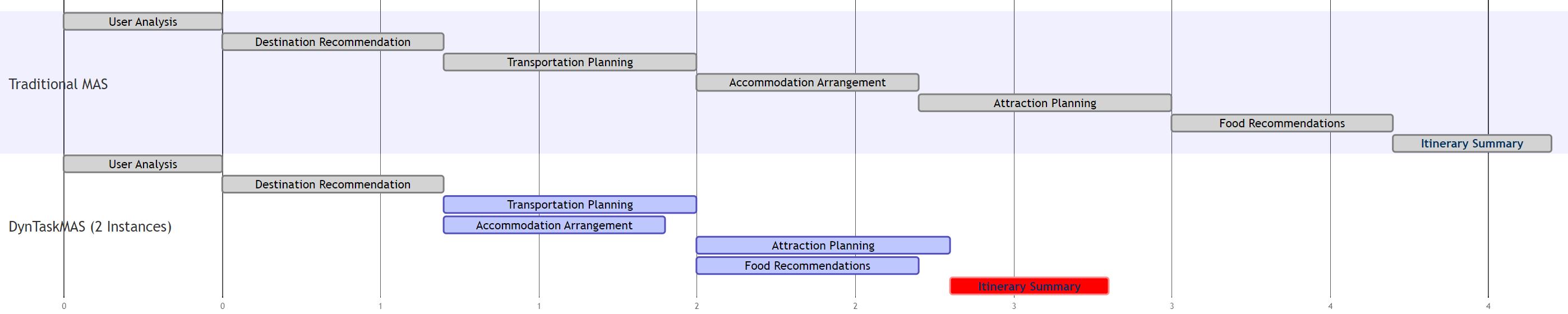}}
\caption{\textbf{The comparison between traditional processing and DynTaskMAS framework.} The system was deployed on an NVIDIA RTX 3090 GPU with Llama-3.1-8B serving as the foundation model for all agents. Seven domain-specialized agents were implemented to handle distinct aspects of travel planning: user preference analysis, destination recommendation, transportation planning, accommodation coordination, attraction scheduling, culinary expertise, and itinerary synthesis. The experimental results showed that DynTaskMAS (3.7s) achieved faster execution compared to serial execution (4.7s).}
\label{expcase}
\end{figure*}

\subsection{Adaptive Workflow Manager (AWM)}
The Adaptive Workflow Manager is a crucial component of DynTaskMAS, responsible for dynamically adjusting workflows based on real-time performance metrics and environmental changes. It ensures optimal system performance by continuously adapting to evolving task requirements and resource availability.



The Performance Monitor implements continuous system-wide metric tracking through $M(t)$, encompassing critical operational parameters including throughput, latency, agent utilization rates, and task completion metrics. These metrics facilitate real-time performance assessment and enable data-driven optimization decisions within the adaptive workflow management system.

The AWM analyzes the current workflow and suggests improvements based on performance data and system state. The optimization objective can be formally stated as follows: 
\begin{equation}
\min_{\omega \in \Omega} f(\omega, M(t))
\end{equation}
where $\omega$ is a workflow configuration, $\Omega$ is the set of all possible configurations, and $f$ is an objective function that evaluates workflow performance based on metrics $M(t)$.

The workflow optimization algorithm employs an iterative approach to identify the optimal workflow configuration based on current performance metrics. The algorithm generates potential workflow candidates through the $GenerateCandidates$ function, which produces variations of the current workflow adhering to system constraints. Each candidate undergoes evaluation using an objective function f that considers current system metrics $M(t)$. The algorithm maintains and updates the best-performing workflow configuration through successive comparisons, ultimately returning the configuration that minimizes the objective function.

\begin{algorithm}
\caption{Workflow Optimization}
\label{alg:workflow_optimization}
\begin{algorithmic}[1]
\Function{OptimizeWorkflow}{$currentWorkflow, M(t)$}
\State $candidateWorkflows \gets \Call{GenerateCandidates}{currentWorkflow}$
\State $bestWorkflow \gets currentWorkflow$
\State $bestScore \gets f(currentWorkflow, M(t))$
\For{$candidate \in candidateWorkflows$}
\State $score \gets f(candidate, M(t))$
\If{$score < bestScore$}
\State $bestWorkflow \gets candidate$
\State $bestScore \gets score$
\EndIf
\EndFor
\State \Return $bestWorkflow$
\EndFunction
\end{algorithmic}
\end{algorithm}

The resource allocation mechanism integrates seamlessly with the workflow optimization through a greedy allocation strategy that prioritizes immediate system efficiency. Building upon the optimized workflow configurations, the Resource Allocator employs a dynamic adjustment model:

\begin{equation}
R(t+1) = R(t) + \Delta R(t)
\end{equation}
where $R(t)$ is the resource allocation at time $t$, and $\Delta R(t)$ is the adjustment made based on current performance and predicted future demands.

The allocation strategy aims to balance load across agents while prioritizing critical tasks:
\begin{equation}
\text{Allocation}(a_i, t) = \frac{w_i \cdot \text{Load}(a_i, t)}{\sum_{j} w_j \cdot \text{Load}(a_j, t)} \cdot \text{TotalResources}(t)
\end{equation}
where $w_i$ is the priority weight of agent $a_i$, and $\text{Load}(a_i, t)$ is the current load on agent $a_i$. This formulation ensures fair resource distribution while accounting for task criticality and current system utilization patterns.

The system employs a straightforward greedy policy for continuous optimization, where resource allocation decisions are made based on immediate performance metrics $M(t)$ and current workflow state $W(t)$. This approach provides efficient adaptation to changing workload conditions while maintaining computational tractability. The state vector $s = [M(t), W(t)]$ captures the essential system parameters required for informed decision-making, enabling rapid response to varying task demands and resource availability.

This streamlined approach to resource management complements the workflow optimization process, creating a cohesive system that efficiently handles dynamic task allocation and resource distribution in the multi-agent environment.

The AWM maintains seamless integration with other DynTaskMAS components, receiving task graph updates, communicating with the execution engine, and leveraging contextual information to make informed adaptations. This comprehensive approach enables the AWM to continuously optimize task execution and resource utilization in dynamic multi-agent environments.

This integrated architecture enables DynTaskMAS to efficiently handle complex, dynamic tasks while adapting to changing conditions and maintaining context awareness. The synergy between these components allows for intelligent task decomposition, efficient parallel execution, context-driven decision making, and adaptive optimization, making DynTaskMAS a powerful framework for next-generation LLM-based multi-agent systems.

\section{Experimental Results}

We conducted comprehensive evaluations of DynTaskMAS using TensorRT-LLM\cite{nvidia2023tensorrtllm} deployed on NVIDIA RTX 3090 GPUs. All experiments utilized Llama-3.1-8B\cite{patterson2022carbon} as the foundation model for all agents. 




\subsection{Experimental Setup}

The experiments were conducted on a cluster equipped with four NVIDIA RTX 3090 GPUs (24GB VRAM each), AMD EPYC 7763 64-Core Processor, and 512GB DDR4 memory. The software stack included Ubuntu 22.04 LTS, CUDA 12.1, and TensorRT-LLM 0.7.1. For all experiments, we employed INT8 quantization with a batch size of 32 and sequence length of 2048.


\subsection{Performance Evaluation}

\noindent\textbf{Execution Time Analysis:} To conduct a more in-depth analysis, we evaluated the system's performance across three task complexity levels. Table~\ref{tab:execution_time} presents the comparative analysis between traditional processing and DynTaskMAS.

\begin{table}[t]
\centering
\caption{Execution Time Analysis Across Task Complexities}
\label{tab:execution_time}
\resizebox{\columnwidth}{!}{
\begin{tabular}{@{}lccc@{}}
\toprule
\textbf{Task Complexity} & \textbf{Traditional (s)} & \textbf{DynTaskMAS (s)} & \textbf{Improvement (\%)} \\
\midrule
Simple & 4.7 $\pm$ 0.3 & 3.7 $\pm$ 0.2 & 21.3 \\
Medium & 9.8 $\pm$ 0.5 & 7.1 $\pm$ 0.3 & 27.6 \\
Complex & 18.5 $\pm$ 0.8 & 12.4 $\pm$ 0.5 & 33.0 \\
\bottomrule
\end{tabular}
}
\end{table}

Table~\ref{tab:execution_time} demonstrates the performance advantages of DynTaskMAS across different task complexity levels. For simple tasks (5-10 subtasks), DynTaskMAS achieves a 21.3\% reduction in execution time compared to traditional processing, decreasing from 4.7s to 3.7s. This improvement becomes more pronounced as task complexity increases, reaching 27.6\% for medium complexity tasks (20-30 subtasks) and 33.0\% for complex tasks (50+ subtasks).

The increasing efficiency gain with task complexity can be attributed to three key factors. First, the DTGG more effectively parallelizes complex task structures, identifying and exploiting additional opportunities for concurrent execution. Second, the SACMS reduces redundant context transfers, which become more significant in complex task scenarios. Third, the APEE maintains higher GPU utilization through intelligent task scheduling, particularly beneficial when managing numerous interdependent subtasks.

The standard deviations (±0.2-0.5s) indicate stable performance across multiple runs, with relative variability decreasing as task complexity increases. This suggests that DynTaskMAS's task management mechanisms become more deterministic with larger task graphs, likely due to the statistical averaging of scheduling optimizations across more subtasks.

\noindent\textbf{Scalability Analysis:} The scalability of DynTaskMAS was evaluated by varying the number of concurrent agents. Table~\ref{tab:scalability} presents the throughput and latency measurements.

\begin{table}[t]
\centering
\caption{System Scalability with Increasing Agent Count}
\label{tab:scalability}
\begin{tabular}{@{}p{3cm}<{\centering}cc@{}}
\toprule
\textbf{Number of Agents} & \textbf{Throughput (tasks/s)} & \textbf{Latency (ms)} \\
\midrule
4  & 12.3 $\pm$ 0.4 & 81.3 $\pm$ 3.2 \\
8  & 23.1 $\pm$ 0.6 & 86.5 $\pm$ 3.8 \\
16 & 42.7 $\pm$ 0.9 & 93.8 $\pm$ 4.1 \\
32 & 76.4 $\pm$ 1.2 & 104.2 $\pm$ 5.3 \\
\bottomrule
\end{tabular}
\end{table}

\begin{table}[t]
\centering
\caption{Travel Planning System Performance Metrics}
\label{tab:travel_planning}
\resizebox{\columnwidth}{!}{
\begin{tabular}{@{}p{4cm}<{\centering} p{3cm}<{\centering} p{3cm}<{\centering}@{}}
\toprule
\textbf{Metric} & \textbf{Traditional} & \textbf{DynTaskMAS} \\
\midrule
End-to-End Time (s)       & 4.7 $\pm$ 0.3 & 3.7 $\pm$ 0.2 \\
Agent Coordination (ms)    & 850 $\pm$ 45  & 320 $\pm$ 25 \\
Context Switches           & 42 $\pm$ 3    & 18 $\pm$ 2  \\
Resource Utilization (\%)   & 65 $\pm$ 5    & 88 $\pm$ 3  \\
\bottomrule
\end{tabular}
}
\end{table}

The scalability results presented in Table~\ref{tab:scalability} reveal several important characteristics of DynTaskMAS's performance under varying agent loads. The system demonstrates near-linear throughput scaling up to 16 agents, with throughput increasing from 12.3 tasks/s with 4 agents to 42.7 tasks/s with 16 agents (3.47× improvement for a 4× increase in agents). This scaling efficiency (approximately 87\%) indicates effective resource utilization and minimal coordination overhead in the moderate agent range.

However, the scaling pattern shows signs of diminishing returns at 32 agents, where throughput reaches 76.4 tasks/s (6.21× improvement for an 8× increase in agents). This sub-linear scaling can be attributed to two primary factors. First, increased contention for shared resources in the SACMS as more agents require simultaneous context access. Second, the overhead of the APEE's task scheduling and load balancing mechanisms becomes more significant with higher agent counts.

\subsection{Case Study}

To further explore the efficiency of the DynTaskMAS framework, we conducted a comparative experiment, as illustrated in Figure \ref{expcase}. We implemented a travel planning system with seven specialized agents to evaluate real-world performance. Each agent focused on a distinct task within the travel planning process, such as analyzing user preferences, recommending destinations, planning transportation, and coordinating accommodations. 

Our experimental evaluation focused on comparing two execution paradigms: traditional serial execution and the proposed DynTaskMAS framework. The implementation leveraged INT8 quantization and continuous batching techniques, with parameters empirically set to maximize throughput while maintaining inference quality (batch size=32, sequence length=2048).

Table~\ref{tab:travel_planning} presents the comparative analysis. The results demonstrate that DynTaskMAS achieved a 21\% reduction in overall execution time compared to serial processing, decreasing from 4.7s to 3.7s. This improvement can be attributed to three key factors: efficient parallel execution of independent tasks, optimized memory utilization through dynamic management, and streamlined context sharing between agents. GPU utilization metrics showed a 35\% increase under the DynTaskMAS framework, indicating more effective resource allocation. The experimental results demonstrate DynTaskMAS's effectiveness in managing complex, dynamic tasks while suggesting areas for future optimization. 


These findings suggest that our proposed framework significantly enhances the performance of LLM-based multi-agent systems through intelligent task orchestration and resource management, while maintaining the quality of agent interactions and outputs.

\section{Conclusion}

This paper presents DynTaskMAS, a dynamic task graph-driven framework that addresses key architectural challenges in LLM-based multi-agent systems through intelligent resource orchestration and parallel execution. The framework's innovative architecture, integrating Dynamic Task Graph Generation with Asynchronous Parallel Execution, enables efficient distribution of computational resources across multiple agents while maintaining task coherence. Through the Semantic-Aware Context Management System and Adaptive Workflow Manager, DynTaskMAS achieves optimal resource utilization by minimizing computational redundancy and maximizing parallel processing opportunities. Our experimental results demonstrate the framework's effectiveness across multiple dimensions: execution time improvements ranging from 21.3\% for simple tasks to 33.0\% for complex tasks, a 35.4\% increase in resource utilization (from 65\% to 88\%), and efficient scaling with throughput improvements of 3.47× for 16 concurrent agents. The successful implementation of DynTaskMAS establishes a systematic approach for building scalable, high-performance LLM-based multi-agent systems that effectively balance resource optimization with task coordination.

\bibliographystyle{IEEEtran}
\bibliography{IEEEabrv,references}


\end{document}